# Impact of Cation Stoichiometry on the Crystalline Structure and Superconductivity in Nickelates

**Yueying Li[1,2], Wenjie Sun[1,2], Jiangfeng Yang[1,2], Xiangbin Cai[3], Wei Guo[1,2], Zhengbin Gu[1,2], Ye Zhu[4] and Yuefeng Nie[1,2*]**

[1] National Laboratory of Solid State Microstructures, Jiangsu Key Laboratory of Artificial Functional Materials, College of Engineering and Applied Sciences, Nanjing University, Nanjing, China
[2] Collaborative Innovation Center of Advanced Microstructures, Nanjing University, Nanjing, China
[3] Department of Physics, The Hong Kong University of Science and Technology, Hong Kong, China
[4] Department of Applied Physics, Research Institute for Smart Energy, The Hong Kong Polytechnic University, Hong Kong, China

\* Correspondence:
Yuefeng Nie
ynie@nju.edu.cn

**Keywords: nickelate film, infinite layer, superconductivity, molecular beam epitaxy, cation stoichiometry**

## ABSTRACT

The recent discovery of superconductivity in infinite-layer nickelate films has aroused great interest since it provides a new platform to explore the mechanism of high-temperature superconductivity. However, superconductivity only appears in the thin film form and synthesizing superconducting nickelate films is extremely challenging, limiting the in-depth studies on this compound. Here, we explore the critical parameters in the growth of high quality nickelate films using molecular beam epitaxy (MBE). We found that stoichiometry is crucial in optimizing the crystalline structure and realizing superconductivity in nickelate films. In precursor $NdNiO_3$ films, optimal stoichiometry of cations yields the most compact lattice while off-stoichiometry of cations causes obvious lattice expansion, influencing the subsequent topotactic reduction and the emergence of superconductivity in infinite-layer nickelates. Surprisingly, *in-situ* reflection high energy electron diffraction (RHEED) indicates that some impurity phases always appear once Sr ions are doped into $NdNiO_3$ although the X-ray diffraction (XRD) data are of high quality. While these impurity phases do not seem to suppress the superconductivity, their impacts on the electronic and magnetic structure deserve further studies. Our work demonstrates and highlights the significance of cation stoichiometry in superconducting nickelate family.

# INTRODUCTION

Over the past decades, there have been a great number of investigations on the superconductivity in nickelates, as they are natural analogs of high-$T_c$ cuprates [1-8]. More recently, superconductivity was eventually found in the hole-doped infinite-layer nickelates [9], which have similar layered structure and $3d^{9-x}$ electronic configuration to those of cuprate superconductors. This significant discovery provides a new platform to explore the mechanism of high-temperature superconductivity and triggers intense research interests[10-16]. Comprehensive theoretical studies have been reported [17-25], whereas experimental progress is still limited and several key issues remain unsolved. First, nickelates have displayed some distinct properties from cuprates despite similar structures [13, 22, 26]. Second, superconductivity has only been observed in nickelate thin films while bulk samples show insulating behavior [27, 28]. These puzzles cast shadow on the understanding of underlying physics of high-$T_c$ superconductivity. Therefore, more experimental progress is undoubtedly required.

However, the difficulty to reproduce the superconductivity in infinite-layer nickelates is obvious in light of only a few precedents for successful synthesis of superconducting nickelate [9, 12, 29, 30]. Recent reports of the observation on the superconductivity in hole-doped LaNiO$_2$, which was not superconducting previously, also emphasize the importance of the film quality [31, 32]. Some influential factors are reported, for instance, the increasing target ablation, different laser fluences in the pulsed laser deposition (PLD) and (002) peak positions in XRD scans, offering meaningful guidance for the Nd$_{1-x}$Sr$_x$NiO$_3$ growth [33]. Some indirect evidence hints their relevance to the stoichiometry [34, 35], which deserves, but still lacks, a thorough investigation.

In this work, we employed MBE to grow perovskite neodymium nickelate films with different cation stoichiometries which are of significance in optimization and reproduction of superconductivity in nickelate films. We found that off-stoichiometry in both nickel-rich and nickel-poor leads to obvious lattice expansion, which is shown to hinder the subsequent topotactic reduction and the emergence of superconductivity in infinite-layer nickelates. Additionally, based on the stoichiometry effect, out-of-plane (OOP) lattice constant is found to be helpful in the MBE growth calibration. Finally, an impurity phase in the Sr-doped samples was always shown in RHEED patterns, which can coexist with superconductivity.

# METHODS

The NdNiO$_3$ and Nd$_{1-x}$Sr$_x$NiO$_3$ films were epitaxially grown on TiO$_2$-terminated (001)-oriented SrTiO$_3$ single-crystalline substrates using a DCA R450 MBE system. Before the growth, we used the quartz crystal microbalance (QCM) to measure a rough beam flux. The value of flux is for reference only, for it depends strongly on the background pressure, installation angle of the crucibles of sources and the shape of the source materials. During the growth, RHEED was employed to monitor the growth process and surface quality. The films were grown at 550 – 650 °C (measured by thermocouple thermometer) and under an oxidant (distilled ozone) background pressure of ~ $4.0 \times 10^{-6}$ Torr. Residual gas analyzer (RGA) was utilized to real-time monitor the ozone partial pressure which is essential for the stabilization of the oxidation state of $Ni^{3+x}$. $SrTiO_3$ substrates were etched in buffered HF acid for about 70 s and annealed in flowing pure oxygen at 1000 °C for 80 min before growth to obtain $TiO_2$-terminated step-and-terrace surface [36]. The film crystalline structure was examined by XRD using a Bruker D8 Discover diffractometer. The terraced micromorphology of films was revealed by Asylum Research MFP-3D atomic force microscopy (AFM). Specimens for the cross-sectional scanning transmission electron microscopy (STEM) were prepared by focused ion beam (FIB) techniques. Atomic-resolution annular dark-field (ADF) images were acquired on the JEOL JEM ARM 200F outfitted with an ASCOR fifth-order probe corrector. In order to attain infinite-layer phase, the $Nd_{0.8}Sr_{0.2}NiO_3$ was sealed in a vacuum chamber together with ~0.1 g $CaH_2$ powder, and then heated to 280 °C for 4h, with warming (cooling) rate of 10-15 °C /min [9]. Temperature-dependent resistivity was measured via the standard Van der Pauw geometry in a homemade transport properties measurement system.

## RESULTS AND DISCUSSION

### 1    Effect of cation stoichiometry on NdNiO$_3$ films

Taking the advantage of MBE technique, a series of $NdNiO_3$ films with different Nd:Ni flux ratio were grown. For many perovskite oxides $ABO_3$, the deposition time for each source can be extracted precisely using a shuttered mode [37, 38], because the alternative growth of $AO$ and $BO_2$ monolayers leads to RHEED intensity oscillations with intensity saturating at the end of depositing one full atomic monolayer. However, the intensity doesn't saturate at the end of growing an atomic monolayer for $NdNiO_3$. Hence, the co-deposition method where $AO$ and $BO_2$ layers are deposited simultaneously is adopted. The period of RHEED oscillations in the co-deposition corresponds to the growth of one unit cell [39, 40]. We adjust the Nd:Ni flux ratio precisely by successively changing the deposition time for Ni, which effectively alters the cation stoichiometry.

XRD patterns shown in **FIGURE 1A** demonstrate clear (*00l*) reflections, indicating the reasonable crystalline quality in these samples. **FIGURE 1B** shows a plot of the OOP lattice constants calculated from (002) peak position as a function of nominal Nd:Ni flux ratio (represented by the blue triangles), as well as the corresponding Gaussian fit (yellow dashed line). An increment of OOP lattice constant is observed in the both sides of flux ratio deviated from the optimal value. Given that the in-plane lattice is fully strained to the substrates as revealed by the reciprocal space mapping (RSM) shown in **FIGURE 1C**, it is clear that off-stoichiometry gives rise to a lattice expansion, similar with the situation in $SrTiO_3$ [41]. Since the (002) peak position over 48° was deemed to be indispensable for superconductivity [33], we note the sensitivity of the OOP lattice constant to cation stoichiometry should attract more attention.

Based on this finding, the OOP lattice constant can be employed in turn as a unique indicator of stoichiometry and aided in the calibration of beam flux ratio which is essential in MBE growth. As mentioned above, the precise deposition time of each source is not available using shuttered mode. In the co-deposition, although the oscillations are observed, the overall intensity shows little dependence on the variation of Nd:Ni flux ratio, which is commonly employed in the growth of other systems [42, 43]. Hence, another specific way is demanded to conduct the calibration. Using OOP lattice constant as an indicator is proved to be feasible and reliable. As shown in **FIGURE 1B**, the *c*-lattice constants as a function of stoichiometry can be nicely fitted with the Gaussian function shown below:

$$y = 3.797 - 0.037 \times e^{-(\frac{x-1.006}{0.083})^2} \quad (1.1)$$

The deviation of flux ratio from optimum can be estimated from the fitted parabolic function. A real practice of the calibration process is specifically shown in **FIGURE 1B** denoted by the red stars. The $NdNiO_3$ samples were grown in order shown by the numbers in the figure. OOP lattice constant of the first sample is indicated by the black dashed line. Thus, the deviation of beam flux ratio away from the optimum can be determined. Note that both No.1 and No.1' positions are possible for this sample with only the lattice constant known. Hence, No.2 and No.3 samples were both grown, in which the dosage of nickel was reduced and increased respectively according to the deviation. The values of OOP lattice constant of the two samples are within expectations and the Nd:Ni flux ratio for No.3 is nearly the optimum. No.4 sample was also grown to further prove the validity of this method and the result was consistent. It should be noted that other factors such as anion concentration also affects the lattice [44], which could explains the slight changes of exact OOP lattice constants of our films.

Then, a series of $NdNiO_3$ films were grown using the calibration process mentioned above. The persistent RHEED oscillations confirm the layer-by-layer growth mode, the period of which marked in **FIGURE 2A** is exactly the time required to deposit a layer of one unit cell $NdNiO_3$. The thickness obtained from RHEED oscillation curve is 27 u.c., in good agreement with the fit of Kiessig fringes [45] shown in **FIGURE 2B**. The rocking curve measurement (**FIGURE 2C**) shows a full width at half-maximum (FWHM) value of 0.017°, indicating a high degree of crystalline perfection. The RHEED patterns taken along [110] and [100] direction are shown as insets of **FIGURE 2A**. The half-order diffractions can be observed, manifesting the existence of $NiO_6$ octahedral rotation [46]. In the atomic-resolution ADF-STEM images shown in **FIGURE 2D** and **E**, the abrupt and straight interface between the $SrTiO_3$ substrates and the $NdNiO_3$ film is observed. The film shows high crystalline quality with well-ordered Nd and Ni atoms (denoted by orange and green circle respectively) forming the perovskite lattice, and no defects such as atomic intermixing and stacking faults are observed. The smooth surface with terraced morphology is achieved and revealed by AFM imaging (**FIGURE 2F).**

## 2 Effect of cation stoichiometry on the emergence of superconductivity

The growth of Sr-doped $NdNiO_3$ is conducted using a similar co-deposition method based on both optimal and off-stoichiometric $NdNiO_3$. Shutter times (deposition time per unit cell) for Nd and Ni are corresponding to single period of RHEED intensity oscillations during NdNiO$_3$ co-deposition growth. The period of SrTiO$_3$ film is also calibrated in advance to obtain precise shutter time of Sr. Take Nd$_{0.8}$Sr$_{0.2}$NiO$_3$ for instance, in the one-unit-cell growth, the shutters of Sr, Nd and Ni are opened together and Sr is closed at 20% shutter time, Nd at 80% and Ni at 100%.

The RHEED intensity oscillations of 18 u.c. thick Nd$_{0.8}$Sr$_{0.2}$NiO$_3$ under the optimal flux ratio is shown in **FIGURE 3A**. The oscillations are not perfectly smooth but still sustained for a long time. The disparity in crystalline quality originating from stoichiometry is obvious from the comparison of 2$\theta$-$\omega$ diffraction patterns of Nd$_{0.8}$Sr$_{0.2}$NiO$_3$ films under different flux ratio (**FIGURE 3B and D**). The (002) diffraction peak of off-stoichiometric Nd$_{0.8}$Sr$_{0.2}$NiO$_3$ film is obviously weaker and broader and the peak position is below 48º marked by the yellow dashed line, let alone its (001) peak nearly indistinguishable. No infinite-layer phase is detectable after the reduction though under the same annealing condition, and the insulating behavior shown in **FIGURE 3C** is also as expected. As such, our results demonstrate the significance of stoichiometry, which will help to optimize the synthesis of nickelate.

Furthermore, for the optimal Nd$_{0.8}$Sr$_{0.2}$NiO$_3$ (inset of **FIGURE 3A)**, though the diffraction pattern of Nd$_{0.8}$Sr$_{0.2}$NiO$_3$ is clear and sharp as shown by the green arrows, there exists impurity phases as indicated by the red dashed open circles, so do most of our Sr-doped samples. It should be noted that no corresponding diffraction peak was detected in XRD scan, suggesting possible short-range order of the impurity phases, but its chemical composition is not clear up to now. Even so, these impurity phases do not seem to suppress the superconductivity in Nd$_{1-x}$Sr$_x$NiO$_2$ (**FIGURE 3C**). As shown in **FIGURE 3B**, after topotactic reduction, the OOP lattice constant shrinks to ~3.38 Å (calculated from (001) peak position). According to the documented doping level dependence of OOP lattice constant [12, 15], the actual Sr concentration in our film is consistent with the nominal value determined in the growth process. We also employed the Scherrer equation (2.1) to estimate the thickness of nickelate film in infinite-layer phase[33, 47]

$$d_{Scherrer} = \frac{k\lambda}{b cos\theta} \quad (2.1)$$

where d$_{Scherrer}$ is the Scherrer thickness, K is the Scherrer constant, 1.091 in our case[33], $\lambda$ is the wavelength of X-ray which is 1.5418 Å, $\theta$ and b are the bragg angle and the full width at half maximum intensity of the corresponding diffraction peak, respectively. The calculated Scherrer thickness (58.07 Å) basically matches with the situation where the precursor perovskite has been fully converted into the infinite-layer structure (60.84 Å). Moreover, XRD patterns of the Nd$_{1-x}$Sr$_x$NiO$_3$ film usually show a double-peak-like feature after capping with SrTiO$_3$ layers, which is reminiscent of the stacking faults in previous reports [33, 48]. However, we measured the same sample before and after capping and found the peak only appearing in the latter (**FIGURE 3B**), implying that the peak at ~48º is more corresponding to the first-order thickness fringe of SrTiO$_3$ capping layer, the intensity of which is enhanced by both the film and substrate.

## CONCLUSION

In summary, optimizing of the quality of nickelate films was investigated in this work using MBE. The crystalline lattice and topotactic reduction of nickelates are both susceptible to off-stoichiometry. Obvious lattice expansion was observed caused by off-stoichiometry in NdNiO$_3$ films, and the crystalline structure as well as transport properties are both influenced after the subsequent topotactic reduction. Our finding is consistent with a previous report, where the (002) peak position over 48º in precursor phase nickelate is deemed as the requisite for superconductivity [33]. In addition, we introduced a new practical method using OOP lattice constant to calibrate the Nd:Ni flux ratio in NdNiO$_3$ growth. Moreover, we found the repetitive appearance of some impurity phases in RHEED patterns for most of our Sr-doped samples, which appear to be unavoidable but do not seem to suppress the superconductivity.

Given the sensitivity of the structure of nickelate films to the variation of cation stoichiometry, any growth parameters that may affect the final stoichiometry in the films should be controlled carefully. For MBE, PLD and many other growth techniques, lots of parameters have impact on the stoichiometry, including beam flux ratio, chemical composition of targets, growth temperature, background pressure, laser plume, laser fluence and target ablation, etc. [34, 35, 49-53]. These growth parameters can be adjusted referring to our findings about the stoichiometry dependence on the OOP lattice constant. Finally, although the superconductivity is not obviously affected by the impurity phases in Sr-doped nickelates, further investigation on their potential impacts on the electronic and magnetic structure is demanded.

## DATA AVAILABILITY STATEMENT

The original contributions presented in the study are included in the article; further inquiries can be directed to the corresponding author.

## AUTHOR CONTRIBUTIONS

YN conceived the project. YL, WS and WG grew the nickelate films. YL, WS, WG and JY conducted the materials and structural characterization. XC and YZ conducted the STEM measurements. JY, YL and WS conducted the reduction experiments. YL and WS performed the transport measurements. YL and YN prepared the manuscript with contribution from all authors. YL acknowledges discussions with JW and HS.

## FUNDING

This work was supported by the National Natural Science Foundation of China (Grant Nos. 11774153, 11861161004, and 51772143), the Fundamental Research Funds for the Central Universities (Grant No. 0213-14380198, 0213-14380167), the Research Grants Council of Hong Kong (N_PolyU531/18) and the Hong Kong Polytechnic University grant (No. ZVRP).

**Non-standard abbreviations used:**
OOP lattice constant: out-of-plane lattice constant

## REFERENCES

1. Chen CH, Cheong S-W, Cooper AS. Charge modulations in $La_{2-x}Sr_xNiO_{4+y}$: Ordering of polarons. *Phys Rev Lett* (1993) 71(15):2461-4. doi: 10.1103/PhysRevLett.71.2461.

2. V. I. Anisimov, D. Bukhvalov, Rice TM. Electronic structure of possible nickelate analogs to the cuprates. *Phys Rev B* (1999) 59(12):6.

3. Lee KW, Pickett WE. Infinite-layer $LaNiO_2$: $Ni^{1+}$ is not $Cu^{2+}$. *Phys Rev B* (2004) 70(16). doi: 10.1103/PhysRevB.70.165109.

4. Chaloupka J, Khaliullin G. Orbital order and possible superconductivity in $LaNiO_3/LaMO_3$ superlattices. *Phys Rev Lett* (2008) 100(1):016404. doi: 10.1103/PhysRevLett.100.016404.

5. Hansmann P, Yang X, Toschi A, Khaliullin G, Andersen OK, Held K. Turning a nickelate Fermi surface into a cupratelike one through heterostructuring. *Phys Rev Lett* (2009) 103(1):016401. doi: 10.1103/PhysRevLett.103.016401.

6. Poltavets VV, Lokshin KA, Nevidomskyy AH, Croft M, Tyson TA, Hadermann J, et al. Bulk magnetic order in a two-dimensional $Ni^{1+}/Ni^{2+}$ ($d^9/d^8$) nickelate, isoelectronic with superconducting cuprates. *Phys Rev Lett* (2010) 104(20):206403. doi: 10.1103/PhysRevLett.104.206403.

7. Han MJ, Wang X, Marianetti CA, Millis AJ. Dynamical mean-field theory of nickelate superlattices. *Phys Rev Lett* (2011) 107(20):206804. doi: 10.1103/PhysRevLett.107.206804.

8. Zhang J, Botana AS, Freeland JW, Phelan D, Zheng H, Pardo V, et al. Large orbital polarization in a metallic square-planar nickelate. *Nat Phys* (2017) 13(9):864-9. doi: 10.1038/nphys4149.

9. Li D, Lee K, Wang BY, Osada M, Crossley S, Lee HR, et al. Superconductivity in an infinite-layer nickelate. *Nature* (2019) 572(7771):624-7. doi: 10.1038/s41586-019-1496-5.

10. Hepting M, Li D, Jia CJ, Lu H, Paris E, Tseng Y, et al. Electronic structure of the parent compound of superconducting infinite-layer nickelates. *Nat Mater* (2020) 19(4):381-5. doi: 10.1038/s41563-019-0585-z.

11. M. Rossi HL, A. Nag, D. Li, M. Osada, K. Lee, B. Y. Wang, S. Agrestini, M. Garcia-Fernandez, Y.-D. Chuang, Z. X. Shen, H. Y. Hwang, B. Moritz, Ke-Jin Zhou, T. P. Devereaux, and W. S. Lee. Orbital and Spin Character of Doped Carriers in Infinite-Layer Nickelates. ArXiv [Preprint] (2020). Available at: https://arxiv.org/abs/2011.00595v1. (Accessed November 1, 2020).

12. Zeng S, Tang CS, Yin X, Li C, Li M, Huang Z, et al. Phase Diagram and Superconducting Dome of Infinite-Layer $Nd_{1-x}Sr_xNiO_2$ Thin Films. *Phys Rev Lett* (2020) 125(14):147003. doi: 10.1103/PhysRevLett.125.147003.

13. Goodge BH, Li D, Lee K, Osada M, Wang BY, Sawatzky GA, et al. Doping evolution of the Mott-Hubbard landscape in infinite-layer nickelates. *Proc Natl Acad Sci U S A* (2021) 118(2). doi: 10.1073/pnas.2007683118.

14. Wang BY, Li D, Goodge BH, Lee K, Osada M, Harvey SP, et al. Isotropic Pauli-limited superconductivity in the infinite-layer nickelate $Nd_{0.775}Sr_{0.225}NiO_2$. *Nat Phys* (2021). doi: 10.1038/s41567-020-01128-5.

15. Li D, Wang BY, Lee K, Harvey SP, Osada M, Goodge BH, et al. Superconducting Dome in $Nd_{1-x}Sr_xNiO_2$ Infinite Layer Films. *Phys Rev Lett* (2020) 125(2):027001. doi: 10.1103/PhysRevLett.125.027001.

16. Gu Q, Li Y, Wan S, Li H, Guo W, Yang H, et al. Single particle tunneling spectrum of superconducting $Nd_{1-x}Sr_xNiO_2$ thin films. *Nat Commun* (2020) 11(1):6027. doi: 10.1038/s41467-020-19908-1.

17. Nomura Y, Hirayama M, Tadano T, Yoshimoto Y, Nakamura K, Arita R. Formation of a two-dimensional single-component correlated electron system and band engineering in the nickelate superconductor $NdNiO_2$. *Phys Rev B* (2019) 100(20). doi: 10.1103/PhysRevB.100.205138.

18. Jiang M, Berciu M, Sawatzky GA. Critical Nature of the Ni Spin State in Doped $NdNiO_2$. *Phys Rev Lett* (2020) 124(20):207004. doi: 10.1103/PhysRevLett.124.207004.

19. Katukuri VM, Bogdanov NA, Weser O, van den Brink J, Alavi A. Electronic correlations and magnetic interactions in infinite-layer $NdNiO_2$. *Phys Rev B* (2020) 102(24). doi: 10.1103/PhysRevB.102.241112.

20. Leonov I, Skornyakov SL, Savrasov SY. Lifshitz transition and frustration of magnetic moments in infinite-layer $NdNiO_2$ upon hole doping. *Phys Rev B* (2020) 101(24). doi: 10.1103/PhysRevB.101.241108.

21. Wu X, Di Sante D, Schwemmer T, Hanke W, Hwang HY, Raghu S, et al. Robust $d_{x^2-y^2}$-wave superconductivity of infinite-layer nickelates. *Phys Rev B* (2020) 101(6). doi: 10.1103/PhysRevB.101.060504.

22. Botana AS, Norman MR. Similarities and Differences between $LaNiO_2$ and $CaCuO_2$ and Implications for Superconductivity. *Phys Rev X* (2020) 10(1). doi: 10.1103/PhysRevX.10.011024.

23. Zhang Y, Lin L-F, Hu W, Moreo A, Dong S, Dagotto E. Similarities and differences between nickelate and cuprate films grown on a $SrTiO_3$ substrate. *Phys Rev B* (2020) 102(19). doi: 10.1103/PhysRevB.102.195117.

24. Sakakibara H, Usui H, Suzuki K, Kotani T, Aoki H, Kuroki K. Model Construction and a Possibility of Cupratelike Pairing in a New $d^9$ Nickelate Superconductor (Nd,Sr)$NiO_2$. *Phys Rev Lett* (2020) 125(7):077003. doi: 10.1103/PhysRevLett.125.077003.

25. Been E, Lee W-S, Hwang HY, Cui Y, Zaanen J, Devereaux T, et al. Electronic Structure Trends Across the Rare-Earth Series in Superconducting Infinite-Layer Nickelates. *Phys Rev X* (2021) 11(1). doi: 10.1103/PhysRevX.11.011050.

26. Zhao D, Zhou YB, Fu Y, Wang L, Zhou XF, Cheng H, et al. Intrinsic Spin Susceptibility and Pseudogaplike Behavior in Infinite-Layer $LaNiO_2$. *Phys Rev Lett* (2021) 126(19). doi: 10.1103/PhysRevLett.126.197001.

27. Wang B-X, Zheng H, Krivyakina E, Chmaissem O, Lopes PP, Lynn JW, et al. Synthesis and characterization of bulk $Nd_{1-x}Sr_xNiO_2$ and $Nd_{1-x}Sr_xNiO_3$. *Phys Rev Mater* (2020) 4(8). doi: 10.1103/PhysRevMaterials.4.084409.

28. Li Q, He C, Si J, Zhu X, Zhang Y, Wen H-H. Absence of superconductivity in bulk $Nd_{1-x}Sr_xNiO_2$ *Communications Materials* (2020) 1(1). doi: 10.1038/s43246-020-0018-1.

29. Qiang Gao, Yuchen Zhao, Xingjiang Zhou, Zhu Z. Preparation of superconducting thin film of infinite-layer nickelate $Nd_{0.8}Sr_{0.2}NiO_2$. ArXiv [Preprint] (2021). (Accessed February 23, 2021).

30. Zhou X-R, Feng Z-X, Qin P-X, Yan H, Wang X-N, Nie P, et al. Negligible oxygen vacancies, low critical current density, electric-field modulation, in-plane anisotropic and high-field transport of a superconducting Nd0.8Sr0.2NiO2/SrTiO3 heterostructure. *Rare Metals* (2021). doi: 10.1007/s12598-021-01768-3.

31. Motoki Osada, Bai Yang Wang, Berit H. Goodge, Shannon P. Harvey, Kyuho Lee, Danfeng Li, et al. Nickelate superconductivity without rare-earth magnetism (La,Sr)$NiO_2$. ArXiv [Preprint]. Available at: https://arxiv.org/abs/2105.13494. (Accessed May 27, 2021).

32. Zeng SW, C. J. Li, L. E. Chow, Y. Cao, Z. T. Zhang, C. S. Tang, et al. Superconductivity in infinite-layer lanthanide nickelates.  [Preprint]. Available at: https://arxiv.org/abs/2105.13492. (Accessed May 27, 2021).

33. Lee K, Goodge BH, Li D, Osada M, Wang BY, Cui Y, et al. Aspects of the synthesis of thin film superconducting infinite-layer nickelates. *APL Materials* (2020) 8(4). doi: 10.1063/5.0005103.

34. Preziosi D, Sander A, Barthélémy A, Bibes M. Reproducibility and off-stoichiometry issues in nickelate thin films grown by pulsed laser deposition. *AIP Advances* (2017) 7(1). doi: 10.1063/1.4975307.

35. Breckenfeld E, Chen Z, Damodaran AR, Martin LW. Effects of nonequilibrium growth, nonstoichiometry, and film orientation on the metal-to-insulator transition in $NdNiO_3$ thin films. *ACS Appl Mater Interfaces* (2014) 6(24):22436-44. doi: 10.1021/am506436s.

36. Masashi Kawasaki KT, Tatsuro Maeda, Ryuta Tsuchiya, Makoto Shinohara, Osamu Ishiyama, Takuzo Yonezawa, Mamoru Yoshimoto and Hideomi Koinuma. Atomic Control of the $SrTiO_3$ Crystal Surface. *Science* (1994) 266(5190):3. doi: 10.1126/science.266.5190.1540.

37. J.H. Haeni, C.D. Theis , Schlom DG. RHEED Intensity Oscillations for the Stoichiometric Growth of $SrTiO_3$ Thin Films by Reactive Molecular Beam Epitaxy. *Journal of Electroceramics* (2000) 4:7.

38. Sun HY, Mao ZW, Zhang TW, Han L, Zhang TT, Cai XB, et al. Chemically specific termination control of oxide interfaces via layer-by-layer mean inner potential engineering. *Nat Commun* (2018) 9(1). doi: 10.1038/s41467-018-04903-4.

39. K. BRITZE GM-E. High energy electron diffraction at Si(001) surfaces. *surface science* (1978) 77:11. doi: 10.1016/0039-6028(78)90166-8.

40. Clarke S, Vvedensky DD. Origin of reflection high-energy electron-diffraction intensity oscillations during molecular-beam epitaxy: A computational modeling approach. *Phys Rev Lett* (1987) 58(21):2235-8. doi: 10.1103/PhysRevLett.58.2235.

41. Brooks CM, Kourkoutis LF, Heeg T, Schubert J, Muller DA, Schlom DG. Growth of homoepitaxial SrTiO$_3$ thin films by molecular-beam epitaxy. *Appl Phys Lett* (2009) 94(16). doi: 10.1063/1.3117365.

42. Zhang TW, Mao ZW, Gu ZB, Nie YF, Pan XQ. An efficient and reliable growth method for epitaxial complex oxide films by molecular beam epitaxy. *Appl Phys Lett* (2017) 111(1). doi: 10.1063/1.4990663.

43. Sun HY, Zhang CC, Song JM, Gu JH, Zhang TW, Zang YP, et al. Epitaxial optimization of atomically smooth Sr$_3$Al$_2$O$_6$ for freestanding perovskite films by molecular beam epitaxy. *Thin Solid Films* (2020) 697. doi: 10.1016/j.tsf.2020.137815.

44. Heo S, Oh C, Son J, Jang HM. Influence of tensile-strain-induced oxygen deficiency on metal-insulator transitions in NdNiO$_{3-\delta}$ epitaxial thin films. *Sci Rep* (2017) 7(1):4681. doi: 10.1038/s41598-017-04884-2.

45. Björck M, Andersson G. GenX: an extensible X-ray reflectivity refinement program utilizing differential evolution. *Journal of Applied Crystallography* (2007) 40(6):1174-8. doi: 10.1107/s0021889807045086.

46. Catalano S, Gibert M, Fowlie J, Iniguez J, Triscone JM, Kreisel J. Rare-earth nickelates RNiO$_3$: thin films and heterostructures. *Rep Prog Phys* (2018) 81(4):046501. doi: 10.1088/1361-6633/aaa37a.

47. Jagodzinski H. H. P. Klug und L. E. Alexander: X-ray Diffraction Procedures for Polycrystalline and Amorphous Materials, 2. Auflage. John Wiley & Sons, New York-Sydney-Toronto 1974, 966 Seiten,. (1975) 79(6):553-. doi: 10.1002/bbpc.19750790622.

48. S. W. Zeng XMY, C. J. Li, C. S. Tang, K. Han, Z. Huang, Y. Cao, L. E. Chow, D. Y. Wan, Z. T. Zhang, Z. S. Lim, C. Z. Diao, P. Yang, A. T. S. Wee, S. J. Pennycook, A. Ariando. Observation of perfect diamagnetism and interfacial effect on the electronic structures in Nd$_{0.8}$Sr$_{0.2}$NiO$_2$ superconducting infinite layers. [Preprint] (2021). Available at: https://arxiv.org/abs/2104.14195. (Accessed April 29, 2021).

49. Seo SSA, Nichols J, Hwang J, Terzic J, Gruenewald JH, Souri M, et al. Selective growth of epitaxial Sr$_2$IrO$_4$ by controlling plume dimensions in pulsed laser deposition. *Appl Phys Lett* (2016) 109(20). doi: 10.1063/1.4967450.

50. Kobayashi K. Effect of growth conditions on stoichiometry in MBE-grown GaAs. *Journal of Vacuum Science & Technology B: Microelectronics and Nanometer Structures* (1985) 3(2). doi: 10.1116/1.583135.

51. schiller S, Beister G, Sieber W. Reactive high rate D.C. sputtering: Deposition rate, stoichiometry and features of TiO$_x$ and TiN$_x$ films with respect to the target mode. *Thin Solid Films* (1983) 111:10. doi: 10.1016/0040-6090(84)90147-0.

52. Selinder TI, Larsson G, Helmersson U, Olsson P, Sundgren JE, Rudner S. Target presputtering effects on stoichiometry and deposition rate of Y-Ba-Cu-O thin films grown by dc magnetron sputtering. *Appl Phys Lett* (1988) 52(22):1907-9. doi: 10.1063/1.99740.

53. Ola N, Martin L, F. FH, Arne K. (2007). "*Rare Earth Oxide Thin Films,*"*Growth of Oxides with Complex Stoichiometry by the ALD Technique, Exemplified by Growth of La$_{1-x}$Ca$_x$MnO$_3$*In: Fanciulli M, Scarel G, editors. *Rare Earth Oxide Thin Films*.(Berlin, Heidelberg): Springer Berlin Heidelberg. p. 87-100.

Figures and captions:

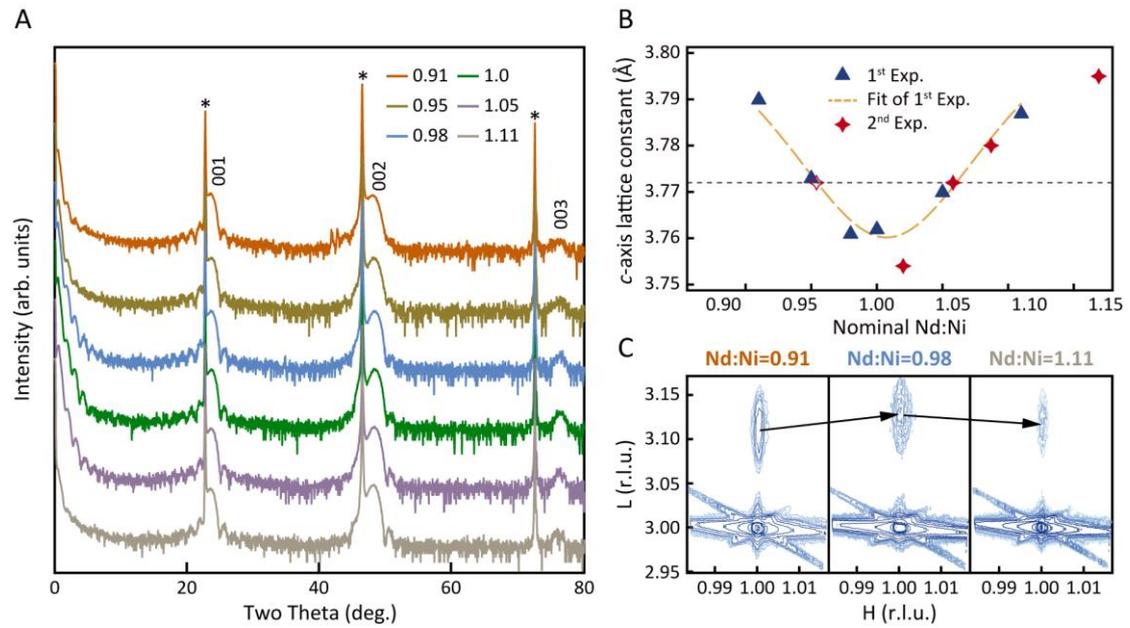

**FIGURE 1| Effect of cation stoichiometry on NdNiO₃ films. A** XRD 2θ-ω scans of NdNiO₃ films with different nominal Nd:Ni flux ratio. The curves are vertically offset for clarity. **B** Pseudo-cubic lattice constant along [*00l*] direction as a function of nominal Nd:Ni flux ratio. The blue triangles and yellow dashed line represent the experimental data and the corresponding Gaussian fit result respectively. The red closed stars denote the data points obtained in another set of experiment, according to the numerical order marked in the figure, while the open star is another possible position for No.1 sample. The black dashed line denotes the *c*-axis lattice constant of the No.1 sample. **C** Reciprocal space mapping of NdNiO₃ films with different flux ratio.

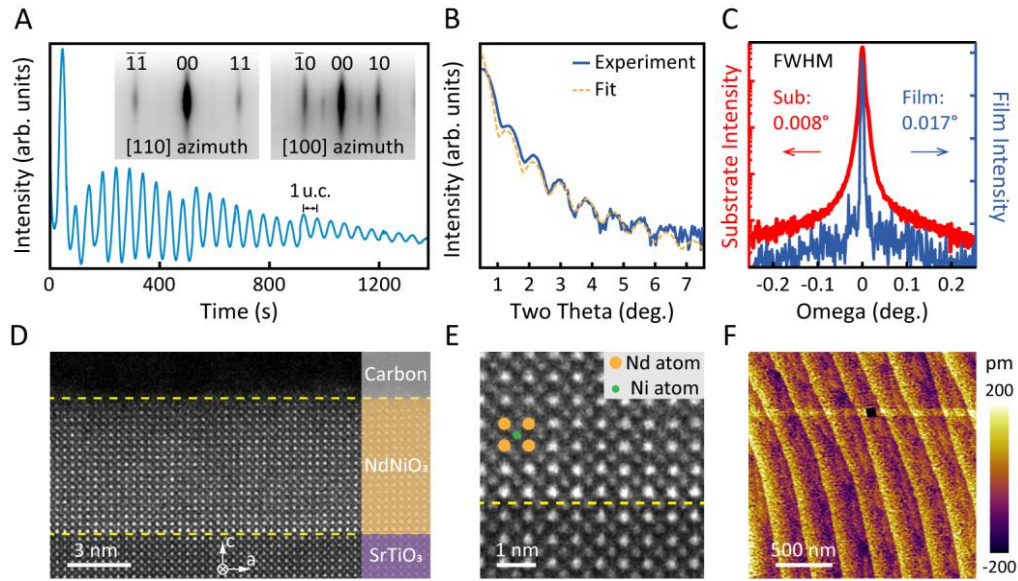

**FIGURE 2| Structural characterizations of optimal NdNiO₃ films. A** RHEED intensity oscillations taken along the [110] direction and RHEED patterns(inset) of a 27 u.c. thick NdNiO₃ film grown on SrTiO₃ substrate. The period of the oscillations denoted by the arrow indicates the time needed for the growth of one-unit-cell-thick NdNiO₃ by co-deposition mode. **B** Magnification of high-resolution XRD 2θ-ω scan of the 27 u.c. NdNiO₃ film at low incident angle with clear

Kiessig fringes and corresponding fit of film thickness shown in yellow dashed line. **C** Rocking curves of the NdNiO$_3$ film and SrTiO$_3$ substrate. **D-E** Cross-sectional STEM-ADF images of an 18-u.c. thick NdNiO$_3$ film on SrTiO$_3$ substrate under different magnification. **F** A representative AFM image of the surface morphology showing clear terraces of NdNiO$_3$ film.

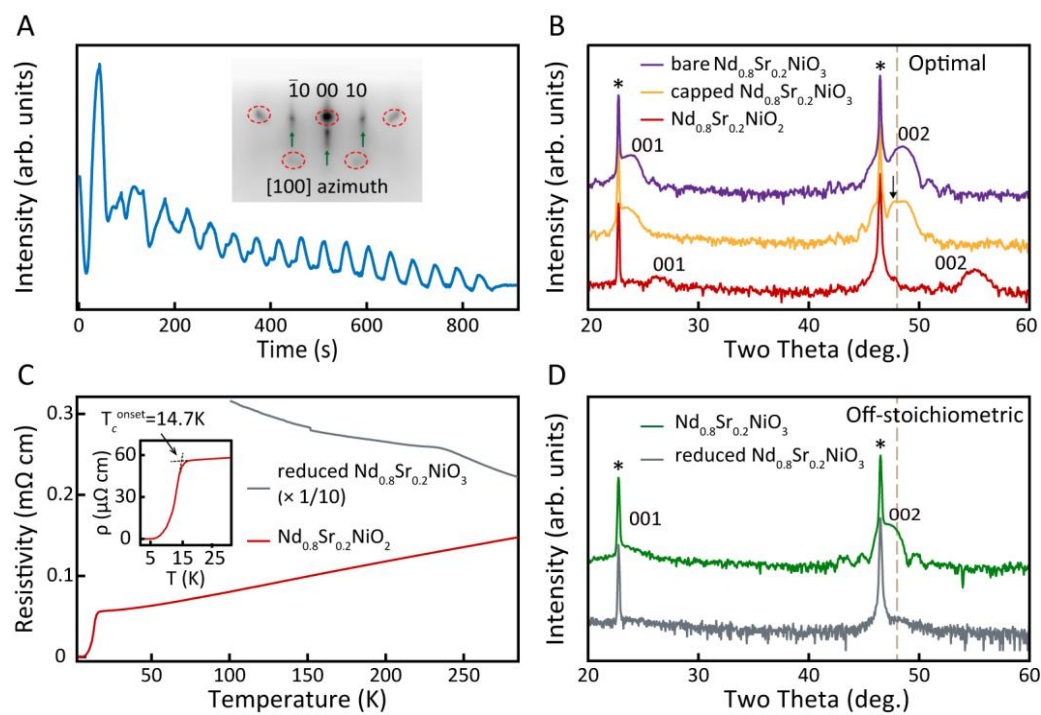

**FIGURE 3| Effect of cation stoichiometry on the reduction of Nd$_{1-x}$Sr$_x$NiO$_3$. A** RHEED intensity oscillations and pattern (inset) of an 18 u.c. thick optimal Nd$_{0.8}$Sr$_{0.2}$NiO$_3$ film grown on SrTiO$_3$ substrate. The red open dashed circles and green arrows indicate the diffractions of the impurity phases and the perovskite phase of the film, respectively. **B** High-resolution XRD 2θ-ω scans of the nickelate films grown under optimal flux ratio. The Nd$_{0.8}$Sr$_{0.2}$NiO$_3$ film before and after capping SrTiO$_3$ layer is called bare and capped Nd$_{0.8}$Sr$_{0.2}$NiO$_3$, respectively. The dashed line represents the critical two theta value of 48°. **C** Temperature dependent resistivity for Nd$_{0.8}$Sr$_{0.2}$NiO$_2$ films grown under optimal and off-stoichiometric flux ratios. The resistivity of reduced Nd$_{0.8}$Sr$_{0.2}$NiO$_2$ was divided by ten times for clarity. Inset shows the zoom-in view at low temperature from 3 K to 30 K. The onset transition temperature is about 14.7 K, and zero resistivity is achieved at about 4.7 K. **D** High-resolution XRD 2θ-ω scans of the nickelate films grown under off-stoichiometric flux ratio. The film after reduction shows no diffraction peaks, thus is called by reduced Nd$_{0.8}$Sr$_{0.2}$NiO$_3$.